\catcode`\@=11

\newif\if@fewtab\@fewtabtrue
{\count255=\time\divide\count255 by 60
\xdef\hourmin{\number\count255}
\multiply\count255 by-60\advance\count255 by\time
\xdef\hourmin{\hourmin:\ifnum\count255<10 0\fi\the\count255}}
\def\ps@draft{\let\@mkboth\@gobbletwo
    \def\@oddhead{}
    \def\@oddfoot
       {\hbox to 7 cm{\tiny \versionno
       \hfil}\hskip -7cm\hfil\rm\thepage \hfil}
    \def\@evenhead{}\let\@evenfoot\@oddfoot}
\def\draftcite#1{\ifnum\draftcontrol=1#1\else{}\fi}
\def\@lbibitem[#1]#2{\item{}\hskip -3cm \hbox to 2cm
{\hfil$\scriptstyle\draftcite{#2}$}\hskip
1cm[\@biblabel{#1}]\if@filesw
     {\def\protect##1{\string ##1\space}\immediate
      \write\@auxout{\string\bibcite{#2}{#1}}}\fi\ignorespaces}
\def\@bibitem#1{\item\hskip -3cm \hbox to 2cm
{\hfil {\footnotesize\draftcite{#1}}}\hskip 1cm
\if@filesw \immediate\write\@auxout
       {\string\bibcite{#1}{\the\value{\@listctr}}}\fi\ignorespaces}
\def\nsection#1{\addtocounter{section}{1} 
\bigskip {\thesection. \ {\large #1} \medskip}}
\def\yes{yes }
 \read-1 to\yesno
\ifx\yesno\yes 
\font\tendl=msym10  scaled \magstep1
\font\sevendl=msym7 scaled \magstep1
\font\fivedl=msym5 scaled \magstep1
\newfam\dlfam \def\dl{\fam\dlfam\tendl}
\textfont\dlfam=\tendl \scriptfont\dlfam=\sevendl
\scriptscriptfont\dlfam=\fivedl
\else \let\dl=\bf
\fi

\global\def\draftcontrol{0}
\catcode`\@=12
\documentstyle[12pt]{article}

\ifx\yesno\yes 
\else  \fi
\def\ifundefined#1{\expandafter\ifx\csname#1\endcsname\relax}
\makeatletter \ifundefined{new@mathgroup} {} \else   
    \new@mathgroup\sffam
    \new@internalmathalphabet\mathsf\sffam{cmss}{m}{n}
    \def\psf{\fontfamily\sfdefault \fontseries\default@series
        \fontshape\default@shape\selectfont\mathsf}
\fi

\def\citen#1{\if@filesw \immediate\write \@auxout {\string\citation{#1}}\fi%
\@tempcntb\m@ne \let\@h@ld\relax \def\@citea{}%
\@for \@citeb:=#1\do {\@ifundefined {b@\@citeb}%
    {\@h@ld\@citea\@tempcntb\m@ne{\bf ?}%
    \@warning {Citation `\@citeb ' on page \thepage \space undefined}}%
    {\@tempcnta\@tempcntb \advance\@tempcnta\@ne
    \setbox\z@\hbox\bgroup\ifcat0\csname b@\@citeb \endcsname \relax
    \egroup \@tempcntb\number\csname b@\@citeb \endcsname \relax
    \else \egroup \@tempcntb\m@ne \fi \ifnum\@tempcnta=\@tempcntb
    \ifx\@h@ld\relax \edef \@h@ld{\@citea\csname b@\@citeb\endcsname}%
    \else \edef\@h@ld{\hbox{--}\penalty\@highpenalty
    \csname b@\@citeb\endcsname}\fi
    \else \@h@ld\@citea\csname b@\@citeb \endcsname \let\@h@ld\relax \fi}%
\def\@citea{,\penalty\@highpenalty\hskip.13em plus.13em minus.13em}}\@h@ld}
\def\@citex[#1]#2{\@cite{\citen{#2}}{#1}}%
\def\@cite#1#2{\leavevmode\unskip\ifnum\lastpenalty=\z@\penalty\@highpenalty\fi%
  \ [{\multiply\@highpenalty 3 #1%
  \if@tempswa,\penalty\@highpenalty\ #2\fi}]}   %
\makeatother 

\def\A             {Algebra}

\newcommand{\authoretc}[5]{\centerline{\sc #1}\vskip2 mm
                   \centerline{#2}\vskip.5mm \centerline{#3}\vskip.5mm
                   \centerline{#4}\vskip.5mm \centerline{#5}}

\def\be            {\begin{equation}}

\newcommand{\Caln}[1]{\mbox{${\cal N}_{\!#1}^{}$}}
\newcommand{\calN}[3]{\mbox{${\cal N}_{\!#1#2}^{\ #3}$}}
\newcommand{\caLN}[4]{\mbox{${}_{#4}^{}{\cal N}_{\!#1#2}^{\ #3}$}}

\def\cft           {conformal field theory}
\def\cfts          {conformal field theories}

\long\def\del#1    \enddel{}

\long\def\drac#1#2{{\displaystyle\frac{#1}{#2}}}

\def\ee            {\end{equation}}
\def\eE            {{\rm e}}

\def\epss          {\mbox{$\epsilon_\sigma$}}
\def\epssm         {\mbox{$\epsilon_{\sigma^{-1}}^{}$}}
\newcommand{\epsS}[1]{\mbox{$\epsilon_\sigma(#1)$}}
\newcommand{\erf}[1]{(\ref{#1})}

\def\findim        {finite-dimensional}

\newcommand{\fline}[1]{\vskip 4mm\noindent ------------------\\[1 mm]}
\def\futnote#1     {\footnote{~#1}\ }

\def\gagr          {Galois group}
\def\Gal           {{\cal G}\!a\ell}
\def\gallq         {\Gal(L/\rationals)}
\def\galml         {\Gal(M/L)}
\def\galmq         {\Gal(M/\rationals)}

\def\gv            {g_{}^\vee}

\newcommand{\hsp}[1] {\mbox{\hspace{#1 em}}}

\def\hy            {$\mbox{-\hspace{-.66 mm}-}$}

\def\id            {{\sl id}}
\def\ii            {{\rm i}}

\def\kma           {Kac\hy Moo\-dy algebra}

\long\def\labl#1   {\label{#1}\ee}

\def\moms          {modular matrix $S$}

\def\mydollar      {$^\pounds$} 

\def\N             {{\dot N}}
\def\Ng            {M}
\def\one           {\mbox{\small $1\!\!$}1}
\def\onedim        {one-dimensional}

\def\qdim          {quantum dimension}

\long\def\query#1{\hskip 0pt{\vadjust{\everypar={}\small\vtop to 0pt{\hbox{}%
     \vskip -13pt\rlap{\hbox to 48.9pc{\hfil{\vtop{\hsize=8pc\tolerance=6000%
     \hfuzz=.5pc\rightskip=0pt plus 3em\noindent#1}}}}\vss}}}}%

\def\qzn           {\rationals(\mbox{$\zeta_n$})}

\def\rationals     {{\dl Q}}

\def\rep           {representation}
\def\Rep           {Representation}

\def\sigmA         {\dot\sigma}
\def\sigmal        {\sigma_{(\ell)}^{}}
\def\sigmAl        {\dot\sigma_{(\ell)}^{}}
\def\sigmalt       {\tilde\sigma_{(\ell)}^{}}
\def\sigmaL        {\sigma_L^{}}
\def\sigmaM        {\sigma_M^{}}

\def\smat          {$S$-matrix}

\newcommand{\sumi}[1] {\mbox{$\displaystyle\sum_{#1\in I}$}}
\newcommand{\sumI}[1] {\mbox{$\sum_{#1\in I}$}}
\newcommand{\sumn}[1] {\mbox{$\displaystyle\sum_{#1=0}^{N-1}$}}

\def\wzwt          {WZW theory}
\def\wzwts         {WZW theories}

\def\9405153       {9405153}
\def\Z             {\mbox{$Z$}}
\def\zet           {{\dl Z}}

\def\zett          {\mbox{\small {\dl Z}}}
\def\zettpluso     {\mbox{${\zett}_{\geq 0}$}}

   \newcommand{\wb}{\,\linebreak[0]} 
   
   \newcommand{\J}[1]     {{{#1}}\vyp}

   \newcommand{\Bi}[1]    {\bibitem{#1}}

   \newcommand{\BOOK}[4]  {{\em #1\/} ({#2}, {#3} {#4})}
   
   \newcommand{\vyp}[4]   {\ {#1} ({#2}) {#3}} 
   \newcommand{\vypf}[5]  {\ {#1} [FS{#2}] ({#3}) {#4}}
   \def\comp  {Com\-mun.\wb Math.\wb Phys.}
   
   \def\ijmp  {Int.\wb J.\wb Mod.\wb Phys.\ A}
   \def\npbf  {Nucl.\wb Phys.\ B\vypf}
   
   \def\nupb  {Nucl.\wb Phys.\ B}
   \def\phlb  {Phys.\wb Lett.\ B}

   \def\phrd  {Phys.\wb Rev.\ D}
   
   \def\WI     {{Wiley Interscience}}
   \def\NY     {{New York}}

\begin{document}

\begin{flushright}  {~} \\[-26 mm] {\sf hep-th/9405153} \\
{\sf IMAFF 94/1}  \\ {\sf NIKHEF-H/94-18} \\[1 mm]{\sf May 1994}
\end{flushright}

\vskip 5mm
\begin{center}
{\Large{\bf MODULAR INVARIANTS AND}} \vskip 0.3cm
{\Large{\bf FUSION RULE AUTOMORPHISMS}} \vskip 0.3cm
{\Large{\bf FROM GALOIS THEORY}}
\end{center}
\vskip 4mm
 \authoretc{J\"urgen Fuchs \ \mydollar}{NIKHEF-H}{Kruislaan 409}
    {NL -- 1098 SJ~~Amsterdam}{~}
 \authoretc{Beatriz Gato-Rivera}{Instituto de Matem\'aticas y F\'\i sica
    Fundamental}{Serrano 123} {E -- Madrid 28006}{~}
 \authoretc{Bert Schellekens} {NIKHEF-H\,/\,FOM} {Kruislaan 409}
    {NL -- 1098 SJ~~Amsterdam}{~}
 \authoretc{Christoph Schweigert} {NIKHEF-H\,/\,FOM} {Kruislaan 409}
    {NL -- 1098 SJ~~Amsterdam}{~}
\vskip 8 mm

\begin{quote} {\bf Abstract.} \
We show that Galois theory of cyclotomic number fields provides a powerful
tool to construct systematically integer-valued matrices commuting with
the modular matrix $S$, as well as automorphisms of the fusion rules.
Both of these prescriptions allow the construction of modular invariants
and offer new insight in the structure of known exceptional invariants.
\end{quote}
\vfill
{}\fline{} {\small \mydollar~~Heisenberg fellow}

\newpage

\nsection{The Galois group and the modular matrix $S$}

The classification of modular invariants and of fusion rule automorphisms
are among the most challenging problems in \cft. In \cite{dego,coga}
it was observed that Galois theory applied to elements of the \moms\ can
shed some light on these issues. In this note we point out that
these connections can be exploited further so that they can actually be
used to construct fusion rule automorphisms and modular invariants.

Given a rational fusion ring with generators $\phi_i,\ i\in I$ ($I$ some
finite index set), and relations
  $ \phi_i \star \phi_j = \sumI k\, \calN ijk \phi_k  $
with $\calN ijk\in\zettpluso$, there exists
a unitary and symmetric matrix $S$ that diagonalizes the
fusion matrices, i.e.\ the matrices \Caln i with entries
$(\Caln i)^k_j:= \calN ijk$. Together with the matrix $T$ with entries
$T_{ij}=T_i\,\delta_{ij}:= \eE^{2\pi \ii (\Delta_i - c/24)}\delta_{ij}$,\,
$S$ generates a \findim\ representation of the modular group $SL_2(\zet)$.
In particular, $S^2=C\,, \ (ST)^3 = C\,, \ C^2 =\one.$
The charge conjugation matrix $C$, a permutation of order
two, will be written as $C_{ij}^{}=\delta^{}_{i,j^+}$.
By the Verlinde formula \cite{verl2}
  \be  \calN ijk = \sumi\ell\, \frac{S^{}_{i\ell}S^{}_{j\ell}S^*_{k\ell}}
  {S_{0\ell}^{}} \,, \labl{verl}
the eigenvalues of the fusion matrices \Caln i are the
{\em generalized quantum dimensions\/} $S_{ij}/S_{0j}$; here the label
$0=0^+\in I$ corresponds to the unit of the fusion ring (or in terms of
\cft, to the identity primary field, i.e.\ to the vacuum of the theory).
They realize the irreducible representations of the fusion ring, i.e.\ we have
  \be \frac{S_{i\ell}}{S_{0\ell}}\, \frac{S_{j\ell}}{S_{0\ell}} =
  \sumi k\, \calN ijk\, \frac{S_{k\ell}}{S_{0\ell}}  \labl{odim}
for all $\ell\in I$.
The generalized \qdim s $S_{il}/S_{0l}$ are the roots of the
characteristic polynomial $\det(\lambda \one - \Caln i)$,
which is a normalized polynomial with integral coefficients, and hence
they are algebraically integer numbers in some algebraic number field
$L$ over the rational numbers \rationals. The extension $L/\rationals$ is
normal \cite{dego}, and hence (using also the fact that
$\rationals$ has characteristic zero) a
{\em Galois extension}; its {\em\gagr} $\,\Gal(L/\rationals)$
is abelian. It follows \cite{dego} that $L$ is contained in some
cyclotomic field $\rationals(\zeta_n)$, where $\zeta_n$ is a primitive
$n$th root of unity.

Applying an element $\sigmaL \in \Gal(L/\rationals)$ on equation \erf{odim}
and using the fact that the fusion coefficients $\calN ijk$ are integers
and hence invariant under $\sigmaL$,
we learn that the numbers $ \sigmaL(S_{ij}/S_{0j}) \,, $
$i\in I$, again realize a \onedim\ representation of the fusion ring.
As the generalized quantum dimensions exhaust all inequivalent \onedim\
representations of the fusion ring \cite{kawA,CUre}, there must exist some
permutation of the labels $j$ which we denote by $\sigmA$, such that
  \be   \sigmaL(\frac{S_{ij}}{S_{0j}})
  = \frac{S_{i\, \sigmA(j)}}{S_{0\, \sigmA(j)}} \,.\labl d

The field $M$ defined as the extension
of \rationals\ that is generated by all \smat\ elements
extends $L$. The extension $M/\rationals$ is again normal and
has abelian \gagr\ \cite{coga}, so that
$\Gal(M/L)$ is a normal subgroup of $\Gal(M/\rationals)$.
Elementary Galois theory then shows that
  \be  0 \to \galml \stackrel\imath\to \galmq \stackrel r\to \gallq
  \to 0 \,, \ee
with $\imath$ the canonical inclusion and $r$ the restriction map,
is an exact sequence, and hence
  \be \Gal(L/\rationals)\cong \Gal(M/\rationals) \,/\, \Gal(M/L) \,.  \ee
In particular any $\sigmaM \in \Gal(M/\rationals)$,
when restricted to $L$, maps $L$ onto itself and equals some element
$\sigmaL \in \Gal(L/\rationals)$. Conversely, any
$ \sigmaL \in \Gal(L/\rationals)$ can be obtained this way. Therefore
by a slight abuse of notation we will frequently use the abbreviation
$\sigma$ for both $\sigmaM$ and its restriction $\sigmaL$.

Working in the field $M$, it follows from \erf d
that for any $\sigmaL\in \gallq$ there exist signs $\epsilon_\sigma(i)
\in\{\pm1\}$ such that the relation
  \be \sigmaM(S_{ij}) = \epss(i)\cdot S_{\sigmA(i)\,j} \labl{18}
is fulfilled for all $i,j\in I$ \cite{coga}. We note that the Galois group
element $\sigma$ and
the permutation $\sigmA$ of the labels that is induced by $\sigma$
need not necessarily have the same order. However, it is easily seen
(see the remarks around \erf2 below) that
an extra factor of 2 is the only difference that can appear.

In this letter we describe how these observations can be extended in two
directions. First, we show that Galois theory
can be used to construct automorphisms of the fusion rules. Second,
we derive from Galois theory a prescription for the systematic construction
of integral-valued matrices in the commutant of the \moms, and hence
of candidate modular invariants. We describe how this method is
implemented for \wzwts. As it turns out, our general prescription is able
to explain many of the modular invariants that are usually referred to as
`exceptional'.

\nsection{Fusion rule automorphisms}

We first show that, if the permutation $\sigmA$ induced by the
Galois group element $\sigma$ leaves the identity fixed,
  \be  \sigmA(0)=0 \,, \labl0
then $\sigmA$ is an automorphism of the
fusion rules. To prove this, we first calculate
  \be \frac{S_{0i}}{S_{00}} = \sigmaL(\frac{S_{0i}}{S_{00}}) =
  \frac{\sigmaM(S_{0i})}{\sigmaM(S_{00})} =
  \frac{\epsilon_\sigma(i)\, S_{0\,\sigmA(i)}}{\epsilon_\sigma(0)\, S_{00}}\, .
  \ee
Since $S_{0j}/S_{00}$, the main (i.e., zeroth) quantum dimensions, are
positive, we learn that the sign $\epss(i)$ is the same for all $i\in I$,
  \be \epsilon_\sigma(i) = \epss(0) =: \epss  = {\rm const}  \,. \labl,
Applying $\sigma$ on the Verlinde formula \erf{verl}, we then find
  \be  \calN ijk =  \sigma( \calN ijk) = \sumi l \frac{\epss^{\!\!3}\,
  S^{}_{\sigmA(i)\,l}S^{}_{\sigmA(j)\,l}S_{\sigmA(k)\,l}^*} {\epss\, S_{0l}}
  = \calN{\sigmA(i)}{\,\sigmA(j)}{\ \ \ \sigmA(k)} \,. \labl{a2}
Next we note that in terms of the cyclotomic field $\qzn\supseteq
M\supseteq L$, the elements $\sigmal \in\gallq$ are simply the restrictions of
elements $\sigmalt\in\Gal(\qzn/\rationals)$; the latter act as
$\zeta_n \mapsto (\zeta_n)^\ell$, and $\Gal(\qzn/\rationals)\cong\zett_n^*$
is the set of all such maps with $\ell$ coprime to $n$. In particular,
$\ell=-1$
corresponds to complex conjugation; the associated permutation of the
generators of the fusion ring is the charge conjugation $C$.
As the Galois group is abelian,
it follows that $\sigmA$ is compatible with charge conjugation,
  \be  \sigmA(i^+)=(\sigmA(i))^+ \,. \labl{a1}
Together with \erf0, the results \erf{a2} and \erf{a1}
show that, as claimed, $\sigmA$ is an automorphism of the fusion rules.

Our result can be interpreted as follows. Let $G:=\{\sigma\in\gallq \mid
\sigmA(0)=0\}$, and let $L^G$ be the subfield in $L$ that is left fixed
under $G$. The elements of the subgroup $G$ of $\gallq$ leave
the main quantum dimensions invariant, and hence the main
quantum dimensions are already contained in $L^G$. The automorphisms of the
fusion rules that are obtained from the \gagr\ as described above are
thus a manifestation of the fact that the main quantum dimensions
do {\em not} exhaust the field spanned by all generalized quantum dimensions.

The general result is nicely illustrated by the example of complex conjugation.
Suppose that the fusion ring is non-selfconjugate, i.e.\ there is at
least one $i\in I$ such that $i^+\ne i$. Then the \moms\
is complex, and as already mentioned the charge conjugation $C$ which
acts as $i\mapsto i^+$ is induced by $\sigma^{}_C=\sigma_{(-1)}\in\gallq$,
i.e. $i^+=\sigmA^{}_C(i)$.
As the main quantum dimensions are real (which is equivalent to $(0)^+=0$),
$G$ contains at least $\sigma^{}_C$ as a nontrivial
element, and charge conjugation is the corresponding non-trivial automorphism.

As a second illustration, consider the extremal case $G =\Gal(L/Q)$. This
means that all main quantum dimensions are rationals (and, since they are
algebraic integers, in fact even ordinary integers). This situation
is realized e.g.\ for $c=1$ \cfts, both for compactification of the free
boson on a circle and for compactification on those $\zett_2$ orbifolds for
which the number of fields is $m^2+7$ for some $m\in\zett$, as well as for
the $({\rm so}(N^2))_2$ and $({\rm su}(3))_3$ \wzwts.
Consider e.g.\ the theory of a free boson on the circle, with $N\in2\zett$
primary fields. The fusion rules read $p \star q= p+q \bmod N$, and the \moms\
has entries $S_{pq}=\eE^{-2\pi\ii\, p q/N}$.
The permutations induced by the \gagr\ are parametrized by $l$, with $l$
and $N$ coprime, and act like $p \mapsto l p \bmod N$. This is invertible
just because $l$ and $N$ are coprime, and clearly an automorphism.
Thus $G$ is the full Galois group, $G\cong\zett_N^*$.
Analogous considerations hold for the
orbifolds and for the \wzwts\ just mentioned.

Note that a permutation automorphism of generic order $N$ does not
directly lead to a modular invariant since the corresponding permutation
matrix $\Pi_\sigma$ generically does not commute with $S$, but rather obeys
$S^{-1} \Pi^{}_\sigma S = \Pi_\sigma^{-1}$. For $N=2$ (such as e.g.\
charge conjugation), $\Pi_\sigma$ does commute with $S$, and hence provides
a candidate modular invariant.
For being indeed a modular invariant, $\Pi_\sigma$ also has to commute
with the modular matrix $T$; it is not difficult to establish
(see the remarks around \erf E below)
that any automorphism of the fusion rules that fulfills
\erf{18} and commutes with the $T$-matrix has order two.

Sometimes there also exist automorphisms of the fusion rules that
{\em cannot\/} be obtained from elements of the Galois group. This
happens for instance if the \smat\ elements of all fields that are
permuted are rational numbers; in this situation, any element of the
Galois group necessarily leaves these fields fixed, and hence cannot
induce the fusion rule automorphism.

\nsection{The construction of \smat\ invariants}

As an easy consequence of the relation \erf{18} between $S_{\sigmA(i)\,j}$
and $S_{ij}$, it follows that for any matrix \Z\ which satisfies
  \be  [\Z,S]=0\,, \qquad \Z_{ij}\in\zett\ \ \forall\ i,j\in i \,, \labl z
the relation
  $ \Z_{\sigmA(i)\,\sigmA(j)}= \epsilon_\sigma(i) \epsilon_\sigma(j)\,\Z_{ij}$
\,holds \cite{coga}. This leads to a selection rule for
those matrices \Z\ which obey $\Z_{ij} \geq 0$
in addition to \erf z, and which hence provide a candidate modular invariant
$ {\cal Z}(\tau,\bar\tau) = \sum_{i,j} \chi_i^*(\bar\tau) \Z_{ij} \chi_j(\tau)$
for the associated \cft\ (this restriction is a generalization of the
`parity rule' of \cite{gann4} and the `arithmetical symmetry' of \cite{rutw2}).

Here we will go beyond the level of mere selection rules and show that
Galois theory can be used to {\em construct\/} modular invariants.
Let us apply $\sigma^{-1}$ to the relation \erf{18}; then we have
  \be S_{ij} = \sigma^{-1} \sigma(S_{ij}) = \sigma^{-1}( \epsilon_\sigma(i)
  S_{\sigmA(i) \,j} ) =  \epsilon_\sigma(i) \epsilon_{\sigma^{-1}}(j) \,
  S_{\sigmA(i)\, \sigmA^{-1}(j)} \,, \labl{33}
where in the last equality one uses the fact that $\epss(i)= \pm 1$ is
rational and hence fixed under $\sigma$. Using \erf{33} $l$ times, we obtain
  \be  S_{ij}=\epsilon_l(i)\epsilon_{-l}(j)\,S_{\sigmA^l(i)\,\sigmA^{-l}(j)}
  \,, \labl y
where the signs $\epsilon_l(i)\equiv\epsilon^{}_{\sigma^l}(i)\in\{\pm 1\}$
are determined by $\epsilon_1\equiv\epss$ through
  $ \epsilon_l(i) = \prod_{m=0}^{l-1} \epsilon_1(\sigmA^m(i))$.
We will employ the simple result \erf{33}, respectively \erf y,
to show that to any element of the Galois group one can associate a matrix
\Z\ which obeys \erf z.

Before proceeding, we should point out that a relation of the form
\erf y need not necessarily stem from Galois theory.
In the proof we actually use only this relation, but not the information
whether it is derived from Galois theory or not.
 \futnote{This remark applies in fact equally to the considerations
about fusion rule automorphisms above.}
In particular, we need not assume that the signs $\epsilon_l$ are
prescribed by some Galois group element $\sigma$, but only use that
they are determined by the permutation $\sigmA$.
However, Galois theory constitutes the only systematic tool that is
known so far to derive such relations, even though it does not
provide an exhaustive list.
(A situation where the symmetry property \erf y of the \moms\ is
satisfied in the absence of Galois symmetries is provided by mutually
local simple currents \cite{scya6} of order two.)
 \futnote{Considering simple currents of general order would amount to
allow the $\epsilon$'s in \erf y to be arbitrary phases instead
of signs. Unfortunately there are no nontrivial cases with $N>2$
and \erf{modinv} being real-valued.}

Thus assume that $\sigmA$ is a permutation, of order $\N$, of the index
set $I$ of a fusion ring and satisfies a relation of the type \erf y,
and define the integer $N$ to be the order of the associated map
$S_{ij}\mapsto \epss(i)\,S_{\sigmA(i)\,j}$. We can then show that for
any set $\{f_l\mid l=1,2,...\,,N\}\subset\zett$ of integers that satisfy
  \be  f_l = f_{-l} \equiv f_{N-l} \,, \labl f
the matrix $Z$ with integral entries
  \be \Z_{jk} := \sumn l\, f_l\,\epsilon_l(k)\,\delta_{j,\sigmA^l(k)}
  \, \labl{modinv}
commutes with the \moms.
Namely, by direct calculation we have
  \be (S\Z)_{ik} = \sumi j \,\sumn l\, S_{ij}\cdot f_l\, \epsilon_l(k)\,
  \delta_{j,\sigmA^l(k)} = \sumn l\, f_l \epsilon_l(k)\,S_{i\,\sigmA^l(k)}
  \ee
as well as
  \be \begin{array}{l}
  (\Z S)_{ik} = \sumi j \,\sumn l \, f_l \epsilon_l(j) \,\delta_{i,\sigmA^l(j)}
  \cdot S_{jk} = \sumn l \,f_l \epsilon_l(\sigmA^{-l}(i))\, S_{\sigmA^{-l}(i)
  \, k} \\[1.9mm] \hsp{2.9} = \sumn l \, f_l \epsilon_l(\sigmA^{-l}(i)) \cdot
            \epsilon_l(\sigmA^{-l}(i))\epsilon_{-l}(k)\, S_{i\,\sigmA^{-l}(k)}
            = \sumn l \, f_l \epsilon_{-l}(k)\, S_{i\,\sigmA^{-l}(k)} \,,
  \end{array} \labl x
where in the transition to the second line we employed \erf y.
Now one merely has to replace the sum on $l$ in \erf x by one on $-l$
and use \erf f to conclude that indeed $S$ and \Z\ commute.
The terms in the sum of \erf{modinv} correspond to the elements of the
cyclic group that is generated by the element $\sigma$ appearing in
\erf{33}; considering more generally an arbitrary abelian group $G$
whose generators satisfy \erf{33}, one proves that the prescription
\erf{modinv} generalizes to
  \be \Z_{jk} = \sum_{\sigma\in G}\, f_\sigma\,\epss(k)\,\delta_{j,
  \sigmA(k)} \,, \labl{Modinv}
with $f_\sigma$ restricted by
  \be  f^{}_\sigma = f^{}_{\sigma^{-1}} \ee
for all $\sigma\in G$.

Returning to the interpretation in terms of the Galois group,
we note that according to \erf{18} the upper limit $N$ of the summation
in equation \erf{modinv} is precisely the order of the Galois group
element $\sigma$ (in particular, Galois theory provides a relation of the
type \erf y with $-l\equiv N-l$), and recall that this order
need not necessarily coincide with the order of
the permutation $\sigmA$ of the labels that $\sigma$ induces.
However, the following consideration
shows that the distinction between $N$ and $\N$ is actually not very
relevant to applications. First, at most a relative factor of 2 can be present;
namely, since $\sigmA$ is of order $\N$, one has in particular $\sigmA^\N
(0)=0$, which by \erf, implies that the sign $\epsilon^{}_{\sigma^\N}$ is
universal, and hence
  \be  \sigma^{2\N}(S_{ij})=\sigma^\N(\epsilon^{}_{\sigma^\N}\,S_{ij})
  =(\epsilon^{}_{\sigma^\N})^2_{}\,S_{ij}=S_{ij}  \,, \labl2
so that $\sigma^{2\N}=\id\,$ on $M$; thus either $N=\N$ or else
$N=2\N$. Furthermore, for $N=2\N$ the terms in the
formula for \Z\ are easily seen to cancel out pairwise, so that the proposed
invariant is identically zero, and hence the case $N\neq\N$ is
rather uninteresting.

We can make another statement about $\sigmA$ by assuming that
it commutes with the $T$-matrix, $T_{\sigmA(i)} = T_i$.
Applying this property together with the relation \erf{33} to the identity
  \be T_i^{-1} S^{}_{ik}T_k^{-1} = \sumi j\, S_{ij} T_j S_{jk}  \ee
which follows from $(ST)^3=S^2=C$, we obtain
  \be \begin{array}{l}  \epss(i)\epssm(k)\, T_i^{-1} S_{\sigmA(i)\,
  \sigmA^{-1}(k)} T_k^{-1} = \epssm(i)\epssm(k)\,\sumi j\,
  S_{\sigmA^{-1}(i)\,\sigmA(j)} T_{\sigmA(j)}
  S_{\sigmA(j)\,\sigmA^{-1}(k)}  \\ {}\\[-2.1 mm] \hsp{12.8}
  =\epssm(i)\epssm(k)\, T_{\sigmA^{-1}(i)}^{-1} S_{\sigmA^{-1}(i)\,
  \sigmA^{-1}(k)} T_{\sigmA^{-1}(k)}^{-1}  \,. \end{array} \ee
Thus
  \be S_{\sigmA^{-1}(i)\,j} = \epss(i)\epssm(i)\, S_{\sigmA(i)\,j}  \labl E
for all $i,j\in I$. As $S$ is unitary, its rows are linearly independent,
and hence \erf E implies that
$\sigmA(i) = \sigmA^{-1}(i)$ for all $i$, i.e.\ that $\sigmA^2=\id$.
Hence any $\sigma$ that fulfills
\erf{18} and commutes with the $T$-matrix has order two.
(Again, this result is just based on the property \erf{33} of $\sigmA$,
and therefore is valid independently of whether $\sigmA$ comes from a
Galois group element $\sigma$ or not.)
As we will see below, at least for \wzwts\ a kind of converse statement
is also true, namely that any \gagr\ element of order two respects the
$T$-matrix up to possibly minus signs.

Due to the presence of the signs $\epss$,
the invariants \erf{modinv} are generically {\em not\/} positive.
However, at least for order $N=2$ one sometimes gets invariants that
are completely positive and moreover have a non-degenerate vacuum.
The only required property of $\sigma$ is that $\epss(i)$ is universal
for all length-two orbits, while the sign for fixed points is arbitrary.
Fixed points with $\epss(i)=-1$ simply get projected out;
in fact, the latter are the only fields that can be directly projected out.

The kind of invariant that is defined by \erf{modinv}
depends on the vacuum orbit. If the
identity is a fixed point, the signs $\epsilon(i) $ are  all equal to
the same overall sign $\epsilon$, as shown in section 2. Then,
for $N=2$, the choice $f_0=0$ and $f_1 = \epsilon$
in \erf{modinv} immediately gives us a positive matrix $\Z$ that
commutes with $S$ and generates a fusion rule automorphism. If the
vacuum is not fixed, the choice $f_0=0$, $f_1 = \epsilon(0)$ leads to
an invariant with an extended chiral algebra in which at least the
identity block is positive. It follows from unitarity of $S$ that
in such an invariant not all coefficients $f_l\epsilon_l(i)$ can be positive
(otherwise $\Z_{ij} \geq \delta_{ij}$, and hence
  $\Z_{00}= \sumI{i,j}S_{0i} \Z_{ij} S_{0j} \geq \sumI iS_{0i}S_{0i} = 1$,
with equality only if $\Z_{ij}=\delta_{ij}$; this is clearly a contradiction).
The only way to get a positive invariant is then that the negative
signs occur precisely for the fixed point orbits, which are then
projected out. If $N=2$ this is indeed possible.
Note that $T$-invariance still remains to be checked in both cases.

For $N > 2$ it is much harder to get a physical invariant. First of
 all
there must exist orbits that violate $T$-invariance, although
such orbits might be projected
out by the summation in \erf{modinv}. It is in fact easy to see that no
positive
integer
invariant can be obtained from from \erf{modinv} if $N$ is odd, for any choice
of $f_l$. If $N$ is odd,
all coefficients except $f_0$ come in pairs $f_l,\ f_{-l}$. It follows that
$\Z_{jj} = f_0$\,mod\,2 for all $j\in I$, and since $\Z_{00} = 1$ this means
that none of the
fields is projected out. Then the unitarity argument given above shows that
a non-trivial positive invariant cannot exist.
If $N > 2$ and even, hence not a prime, one has to distinguish various
kinds of fixed point orbits. Positive
modular invariants may then well exist, but we will not consider this
more complicated case in this paper.

Let us stress that even if the matrix \erf{modinv} contains
negative entries, or does not commute with $T$,
it can still be relevant for the construction of physical invariants,
because the prescription may be combined with other procedures in such
a manner that the negative contributions cancel out.
For example one may use simple currents to extend the chiral algebra
before employing the Galois transformation, or it may happen that a
certain linear combination with
other known elements of the integer commutant of $S$ is a physical invariant.

\nsection{\wzwts}

In the special case where the fusion ring describes the fusion rules of a
\wzwt\ based on an affine Lie algebra $\hat g$ at level $k$, the \gagr\
is a subgroup of $\zett_{\Ng(k+\gv)}^*$, where $\gv$ is the dual
Coxeter number of the horizontal subalgebra $g$ of $\hat g$ and $\Ng$
is the denominator of the metric on the weight space of $g$.
 \futnote{In \cite{coga} a larger cyclotomic field is used in
order to take care of the overall normalization factor of $S$.
But the permutation
of the primary fields can be read off the generalized quantum
dimensions, which do not depend on  the normalization of $S$. The
correct Galois treatment of the normalization of $S$ leads to an overall
sign, which however is irrelevant for our purpose.}

Let us label the primary fields by their $g$-weight
$\Lambda$ which corresponds to an integrable highest weight of
$\hat g$ at level $k$, and denote by $\rho$
the Weyl vector of $g$. Then a Galois transformation labelled by
$\ell\in\zett_{\Ng(k+\gv)}^*$ acts as the permutation \cite{coga}
  \be  \sigmAl(\Lambda)=w(\ell\cdot(\Lambda+\rho))-\rho \,. \labl w
That is, one first performs a dilatation of the shifted weight
$\Lambda+\rho$ by a factor of $\ell$.
The weight $\tilde\Lambda=\ell(\Lambda+\rho)-\rho$ so obtained
is not necessarily an integrable highest weight at level $k$.
If it is not integrable,
then one has to supplement the dilatation by (the horizontal projection of)
an affine Weyl transformation $w\equiv w_{\ell;\Lambda}$. Note that
$\Lambda+\rho$ is an
integrable weight at level $k+\gv$. Using affine Weyl transformations
$w$ at this level we can rotate $\ell(\Lambda+\rho)$ back to another
integrable weight at level $k+\gv$, which is in fact unique. In general
there is no guarantee that after subtraction of $\rho$ one gets an
integrable weight at level $k$, but it is not hard to see that this
does indeed work simultaneously for all integrable weights if the integer
$\ell$ is coprime with $(k+\gv)$, which indeed follows from the requirement
for \erf w to correspond to an element of the \gagr. Finally, there is a
general formula for the sign $\epsilon_{\sigmal}^{}$, namely
  \be  \epsilon_{\sigmal}^{}(\Lambda)=\eta^{}_\ell\,{\rm sign}
  (w_{\ell;\Lambda}) \,,\ee
i.e.\ the sign is just given by that
of the Weyl transformation $w$, up to an overall sign $\eta$ that
only depends on $\sigmal$ \cite{coga}, but not on the individual
highest weight $\Lambda$.

That it is the shifted weight $\Lambda+\rho$ rather than $\Lambda$ that
is scaled is immediately clear from the Kac\hy Peterson formula for
the \moms. In fact, it is possible to derive formula \erf{y} directly
by scaling the row and column labels of $S$ by $\ell$ and $\ell^{-1}$,
respectively,
using \erf{w}. Galois symmetry is thus not required to derive this formula,
nor is it required to show that \erf{modinv} commutes with $S$.
Galois symmetry has however a general validity and is not restricted
to WZW models.

Substituting \erf{w} into the formula for WZW conformal weights one
easily obtains a condition for $T$-invariance, namely
$(\ell^2-1) = 0$ mod $2\Ng(k+\gv)$ \,(or mod $\Ng(k+\gv)$ if all integers
$\Ng(\Lambda+\rho,\Lambda+\rho)$
are even). 
 Since $\ell$ has an inverse mod $\Ng(k+\gv)$, it follows that
$\ell=\ell^{-1}$ mod $\Ng(k+\gv)$, i.e.\
the order of the transformation must be 2, as we already saw in the
general case.

It is straightforward to find solutions to these conditions, and a little
bit more work to check if the resulting modular invariants are indeed
positive. Without any claim to generality we list here some examples
of known invariants that come out in this way: \\[-1.9em]
\begin{itemize} \addtolength{\itemsep}{-.7 em}
\item[--] First of all we get
some (though not all) simple current invariants. The $D$-type invariants
of $A_1$ at level $4m$ appear for $\ell=4m+1$.
In general integer spin invariants of order 2 simple currents
and automorphism invariants generated by fractional spin simple
currents of odd prime order seem to come out as Galois invariants, but
except for $A_1$ we do not have a general proof.
\item[--] Several chiral algebra extensions corresponding to conformal
embeddings \cite{scwa,babo,boNa} are obtained, for example for
$(A_{2})^{}_{5}$, $(A_{4})^{}_{3}$, $(G_{2})^{}_{3}$, $(G_{2})^{}_{4}$ and
$(F_{4})^{}_{3}$.
\item[--] We also found four extensions by currents of spin higher than 1,
namely for $(A_{9})^{}_{2}$, $(D_{7})^{}_{3}$, $(E_{6})^{}_{4}$ and
$(E_{7})^{}_{3}$. The first two are expected on the basis of
rank-level duality \cite{vers2}, and all four appeared already
in \cite{sche5}.
\item[--] Finally we constructed two pure automorphisms for $(G_{2})^{}_{4}$
and
$(F_{4})^{}_{3}$, which were first found in \cite{vers}.
\item[--] In other cases we could obtain positive invariants
after extending the chiral algebra by simple currents, for example
for $(A_{1})^{}_{28}$, $(A_{2})^{}_{9}$, $(A_{3})^{}_{8}$,
$(C_{3})^{}_{4}$, $(C_{4})^{}_{3}$, $(C_{10})^{}_{1}$.  \end{itemize}
  {~} \\[-1.9em]
In some other cases expected invariants appeared as linear combinations.
A detailed description of these and other examples will be presented
elsewhere.

However, there also exist invariants that cannot be explained by Galois
symmetry. One such example is the $E_7$-type invariant of $(A_{1})^{}_{16}$,
an automorphism built on top of the $D$ invariant. This automorphism
is of the form $\erf{y}$, but it relates $S$-elements that are
rational numbers, and hence transform trivially under Galois
transformations.

\nsection{Discussion}

There are several striking similarities between  Galois symmetries and
simple current symmetries. First of all both are related to general
properties of fusion rings, and not to particular (e.g.\ WZW) models.
Both imply
equalities among certain matrix  elements of $S$ up to signs or phases.
Both symmetries organize the fields of the theory into orbits, whose
length is a divisor of the order $N$ of the symmetry. In both cases one can
give
very simple generic formulas for $S$-invariants, and in both cases the
phenomenon of `fixed points', i.e.\ of orbits whose length is less than $N$,
occurs. In both cases such fixed points can appear with multiplicities
larger than 1 in certain modular invariants in which the chiral
algebra is extended. Note that this kind of structure is empirically
observed in nearly all exceptional  (not simple current generated)
invariants found thus far. However, we believe this is the first time that at
least in
some cases the apparent `orbits' and `fixed points' of exceptional
invariants  are actually related to an underlying discrete symmetry.  This
might in
fact be of some help in the still open problem of resolving fixed points of
exceptional invariants.

There is also an important difference between Galois and simple current
symmetries. In the latter case on can give a general construction of
invariants that are positive and are also $T$-invariant.
For Galois invariants it may well be possible to find a general
criterion for $T$-invariance (as we have done for WZW models), but
positivity appears to be a much more difficult requirement.
Experience with \wzwts\ suggests that positive invariants only occur for
low levels. At higher levels the signs $\epsilon(i)$ are
distributed without any
obvious pattern, and since the number of representations increases rapidly,
positive invariants become less and less likely. This explains the
`exceptionality' of these invariants. Unfortunately it is far from obvious
how to make this statement precise (except perhaps for $g=A_1$), and
furthermore we know already one counter-example: the $D$ invariants of
$A_1$ at levels $4m$
form an infinite series of positive Galois invariants.

There is, however, one set of $S$-invariants that is always positive,
namely those due to a $\zett_2$ Galois symmetry that fixes the vacuum. In
WZW models such invariants (that also commute with $T$) are abundant: this
includes all charge conjugation invariants and also at least some of the
simple current automorphism invariants that were first constructed in
\cite{allz}. Remarkably, very few exceptional ones are known.

Let us also note that
formula \erf{a2} can be generalized to automorphisms $\sigmA$
which change the vacuum, i.e.\ obey $\sigmA(0)\neq0$ (and hence are
not automorphisms of the fusion ring as a {\em unital\/} ring).
In this situation, \erf{a2} gets replaced by
 \be \begin{array}{l}  \calN ijk = \sigma(\calN ijk)
  = \sigma(\sumi l \drac{S^{}_{il}S^{}_{jl}S_{kl}^*}{S_{0l}})
  \\ {} \\[-2.2 mm] \hsp{1.7}
  = \sumi l \drac{\epsS i\epsS j\epsS k\, S^{}_{\sigmA(i)\,l}
    S^{}_{\sigmA(j)\,l}S_{\sigmA(k)\,l}^*} {\epsS0 S_{\sigmA(0)\,l}}
  =\epsS0\epsS i\epsS j\epsS k\; \caLN{\sigmA(i)}{\,\sigmA(j)}
  {\ \ \ \sigmA(k)} {\sigmA(0)} \,, \end{array} \labl{nn}
where $\caLN ijkl\equiv \sumI m\, S^{}_{im}S^{}_{jm}S^*_{km} / S_{lm}.$
Note that the numbers $\caLN ijkl$ are well-defined only if $S_{lm}\ne0$
for all $m\in I$, in which case according to \erf{nn} they are actually
integers; in the present situation this condition is met because
$ S_{\sigmA(0)\,i}=\epsS0\epsS i\,S_{0\,\sigmA(i)}\ne0$ for all $i\in I$.
This result can be interpreted as follows. Allowing also for negative
structure constants, we can introduce a second fusion product
$\star_\sigma$, with structure constants $\caLN ijk{\sigmA(0)}$, on the
same ring $\zett^{|I|}$. Defining
  $  \tilde\phi_i:=\epsS0\epsS i\,\phi_{\sigmA(i)}, $  it follows that
$\tilde\phi_i\star_\sigma\tilde\phi_j=\sumI k\,\calN ijk\,\tilde\phi_k,$
i.e.\ both fusion structures are isomorphic. Some special cases of this
phenomenon have already been noticed in \cite{ehol2}. While our
argument uses symmetries of number fields, in \cite{ehol2} the
representation theory of the modular group is employed; thus our
observation suggests a relation between number fields and modular forms.

\vskip1mm
In this paper we have presented a procedure for constructing
modular invariant partition functions directly from symmetries of the
matrix $S$, without any explicit knowledge of its matrix elements.
This method is valid for all rational \cfts, and not a priori restricted
to WZW models and coset theories, unlike conformal embeddings or
rank-level duality. Previously only two such methods were known, namely
charge conjugation (actually an example of Galois symmetry)
and simple currents, and usually the term `exceptional invariant' was used
to refer to anything else. By providing a third general procedure,
the results of this paper define a new degree of `exceptionality'
for modular invariant partition functions. Invariants satisfying this new
definition of exceptionality do exist; this may be taken as an indication
that still more interesting structure remains to be discovered.

\newpage 

\end{document}